\documentstyle{article}
\begin{document}
\begin{sloppypar}

{\LARGE \centerline{{\bf LSDA+U approximation-based analysis of
the}} \centerline{{\bf  electronic structure of CeFeGe$_3$}} }

\vspace{10mm}
\begin{center}
{\large E. Chigo-Anota$^{\dagger}$\footnote{Corresponding author:
Tel: +52 (222) 2295610, Fax: +52 (222) 2295611}, A.
Flores-Riveros$^{\ddagger}$ and J. F. Rivas-Silva$^{\ddagger}$ }

\vspace{4 mm}
\noindent
$^{\dagger}${\sl Posgrado en Ciencias Qu\'{i}micas-Facultad de
Ciencias Qu\'{i}micas, Benem\'erita Universidad Aut\'onoma de Puebla,
Blvd. 14 Sur 6301, 72570 Puebla, Pue., M\'exico,
e-mail: echigoa@sirio.ifuap.buap.mx}

\vspace{4 mm} \noindent $^{\dagger}${\sl Facultad de
Ingenier\'{i}a Qu\'{i}mica, Benem\'erita Universidad Aut\'onoma de
Puebla, Av. San Claudio y Blvd. 18 Sur , 72570 Puebla, Pue.,
M\'exico.}

\vspace{4 mm}
\noindent
$^{\ddagger}${\sl Instituto de F\'{i}sica ''Luis Rivera Terrazas'',
Benem\'erita Universidad Aut\'onoma de Puebla, Apdo. Postal J-48, 72570
Puebla, Pue., M\'exico, e-mail: rivas@sirio.ifuap.buap.mx}

\end{center}

\vspace{15 mm} \centerline{{\sc Abstract}} \noindent We perform
\emph {ab initio} electronic structure calculations of the
intermetallic compound CeFeGe$_3$ by means of the Tight Binding
Linear Muffin-Tin Orbitals-Atomic Sphere Approximation
(TB-LMTO-ASA) within the Local Spin Density Approximation
containing the so-called Hubbard correction term (LSDA+U$^{SIC}$),
using the Stuttgart's TB (Tight Binding)-LMTO-ASA code in the
framework of the Density Functional Theory (DFT).

\vspace{4 mm} \noindent KEYWORDS: Ab initio calculations, LSDA+U
approximation, intermetallic
compound. \\
PACS:  31.15. Ar, 31.15.Ew, 71.27+a

\newpage
\section{Introduction}

\vspace{4 mm} \noindent The heavy electron compounds refer to
those having a specific heat coefficient $\gamma$ of the order of
Jmol$^{-1}$K$^{-2}$, which is much larger than that of simple
me\-tals (being typically in the range of mJmol$^{-1}$K$^{-2}$).
In addition, this implies that such compounds$-$which have been
given a great deal of attention for long time$-$possess a large
effective mass m$^*$, outweighing several hundreds of times the
free electron mass [1]. They can be classified into two groups:
the concentrated Kondo compounds CK and the intermediate valence
compounds (IV) [2,3] depending on the position of the $4f$ or $5f$
level, relative to the Fermi level. The CK compounds have an
integer valence at temperatures much higher ($T\gg T_k$) than the
Kondo temperature $T_k$, at which, there appears the Kondo effect
(an effect observed in metals with a magnetic impurity). However,
at comparatively low temperatures $(T\ll T_k$) they form a Fermi
liquid state [4] with a reduction of its magnetic moment. On the
other hand, the IV compounds do not possess an integer valence at
room temperature as a result of the strong hybridization between
the $4f$ electrons and the conduction electrons, due to the
anomalous proximity between the $4f$ and the Fermi levels.

\vspace{4 mm}
\noindent
Remarkably, the CeFeGe$_3$ compound here studied [5] apparently gives rise
to two behaviors: A high T$_k$ of the order of 100 K and an integer valence
for Ce at room temperature. This study is motivated by the fact that such
compound is known to present a strong electronic correlation character,
which occurs when the Coulomb repulsion between electrons strongly inhibits
their motion, thus becoming highly localized. Because of this, on compounds
containing lanthanides the normally expected metal behavior$-$cerium ions with
$4f$ electrons or uranium and neptunium ions with $5f$ electrons$-$is not
observed. Some examples are: CeA$_{13}$, CeCuSi$_2$, CeCu$_6$, UBe$_{13}$,
UCd$_{11}$, U$_2$Zn$_{17}$ and NpBe$_{13}$, as well as transition metal
oxides, organic metals and carbon compounds, i.e., carbon nanotubes, etc.

\vspace{4 mm} \noindent In the first stage of the present work
\emph {ab initio} calculations were carried out to investigate the
electronic structure of the intermetallic compound CeFeGe$_3$
(tetragonal structure with spatial group 107) by using a DFT
method [7] in the LMTO-ASA approximation [8], whereas in the
second, an LSDA+U$^{SIC}$ approximation [9] was used, as
implemented in the Stuttgart's TB-LMTO-ASA code version 47 [8].
The density of states (DOS), total and partial, for cerium and
iron, as well as the band structure (BS) are obtained from the
compound geometrical conformation optimized by the CASTEP program
(Cambridge Serial Total Energy Package) which makes use of the
ultrasoft pseudopotentials theory [10]. The Coulomb and exchange
parameters used in the calculations were $\mathrm{U}=5.4$ eV [11]
for cerium and $\mathrm{U}= 2.3$ eV [12] $(J=0.9$ eV) for iron,
just obtained in the literature.

\section{Computational approach.}

\subsection{LMTO-ASA and CASTEP Calculations.}

\vspace{4 mm} \noindent The LMTO-ASA approximation employs the
unitary cell splitting into overlaping Wigner-Seitz (WS) spheres
with a maximum overlap of $15\%$, which is generally considered a
reasonable approximation when having a spherically symmetric
potential within the spheres. In addition, use is made of the ASA
(Atomic Sphere Approximation) condition, i.e., a spherical
approximation containing no zone of free electrons on the
Muffin-Tin structure. For open structures, as the ones here
considered, a set of empty spheres is introduced as a device for
des\-cribing the repulsion potentials in the interstices (between
an atomic and an interstitial sphere an overlap of $20\%$ is
allowed). The total volume of the WS spheres is equal to the unit
cell volume, thus eliminating the interstitial region.

\vspace{4 mm}
\noindent
On the other hand, the Muffin-Tin (MT) orbital is energy dependent with the
following linearized form:
\begin{eqnarray}
\phi_{\wedge}(\vec r) = i^l Y_{lm}(\hat{r})
\left\{
\begin{array}{cll}
\phi_\wedge (E,r) + p_{\wedge}(r/S_R)^l\, & ; & r < S_R\\
                            (S/r)^{l+1}\, & ; & r > S_R
\end{array}
\right \}
\end{eqnarray}
where $\phi_{\wedge} (\vec r)$ is found by numerical solution of the radial
Schr\"odinger equation, $\wedge= RL$. $R$ is the site index, whereas $l$ and
$m$ are the orbital and magnetic quantum numbers of the angular momentum.
The $Y_{lm}$'s refer to spherical harmonics and $S_R$ is the MT radius. The
numerical orbital $\phi_{\wedge}(E,r)$ is augmented inside the sphere by a
renormalized spherical Bessel function,
\begin{eqnarray}
J_{k \wedge}(r) = i^l Y_{lm} \frac{(2l+1)!!}{kS_R)^l} j_l(kr),
\end{eqnarray}
whereas outside the spheres, a renormalized spherical Hankel function is added:
\begin{eqnarray}
H_{k \wedge}(r) = i^lY_{lm} \frac{(kS_R)^{l+1)}}{(2l-1)!!} h_l(kr).
\end{eqnarray}
Here, $h_l=j_l-in_l$ is a linear combination of spherical Bessel and Neumann
functions. Thus, the tail of the MT orbitals at $r>S_R$ is the solution of
the Helmholtz equation with zero kinetic energy. The potential parameters
$p_{\wedge}$ are chosen so as to ensure that the wave function be continuous
and differentiable at the sphere boundary.

\vspace{4 mm} \noindent In this approximation the potential
$V_{xc}$ is either von Barth-Hedin (vBH)[13] or
Vosko-Ceperly-Alder [14] type at the Local Spin Density
Approximation (LSDA) level, whereas at the Generalized Gradient
Spin Approximation (GGS) level$-$which is a functional containing
a density gradient correction$-$the potentials used the
Langreth-Mehl-Hu [15] or Perdew-Wang type [16]. The
exchange-correlation potential used in the present calculations
corresponds to that containing the von Barth-Hedin
parametrization, whose general form is
\begin{eqnarray}
v^{\sigma}_{xc} = A(r_s)
\left(\frac{2n_{\sigma}}{n}\right)^{\frac{1}{3}} + B(r_s),
\end{eqnarray}
where $A(r_s)$ and $B(r_s)$ are analytical functions.

\subsection{Geometry Optimization}

\vspace{4 mm}
\noindent
The ternary system unit cell was optimized
by means of ab initio methods, based on a DFT treatment
within the Local Spin Density Approximation (LSDA) and with the
Generalized Gradient Spin (GGS) approximation. In the CASTEP
program [10] the wave function is expanded in planes waves for the
valence electrons, whereas the core electrons (bound to the
nucleus) are taken into account by means of their effective
interaction on the valence electrons, in the form of
pseudopotentials which are added to the corresponding Kohn-Sham
Hamiltonian.

\vspace{4 mm}
\noindent
The pseudopotentials used in this work
were generated by Vanderbilt [17] in the Kleinman-Bylander [18]
representation. The parametrizations of polarized spin developed
by Perdew-Zunger [19] and Perdew-Burke-Ernzerhof [20] for the
exchange and correlation energy were used. The conjugated gradient
method is employed to relax nuclear positions. The sampling of the
Brillouin zone was of $7 \times 7 \times 8$ and $9 \times 9 \times
11$ for the LSDA and GGS approximations, respectively, using the
scheme of Monkhorst-Pack [21]. The cutoff energy for the plane
waves was of 400 eV approximately. Self-consistency in the
calculations was attained whenever the total energy changes were
$\leq$ 5 meV, which corresponds to a criterion of reasonably good
convergence.

\vspace{4 mm}
\noindent
To perform geometry optimization we let
the lattice constants $a,b$ and $c$ vary as free
parameters$-$though they are expected to undergo changes no larger
than 5\% relative to the experimental values. The crystal energy
is minimized with respect to the degrees of freedom by taking into
account the calculation of Hellmann-Feynman forces in the atoms
and the components of the stress tensor [22]. Finally, the
utilized optimization criteria were 0.00002 eV, $0.0010$ \AA \-
and $0.050$  eV/\AA \- for energy change, quadratic mean
displacement, and quadratic mean force per atom, respectively.

\subsection{Mathematical Structure of the approximation LDA+U$^{SIC}$}

\vspace{4 mm} \noindent In the LSDA+U [9] method the electrons are
separated into two subsystems (i and ii). For Ce, (i) delocalized
$s, p$ and $d$ electrons, which are described by an
orbital-independent one-electron potential (the LSDA potential),
and (ii) localized $4f$ electrons, for which we take into account
the orbital degeneracy and a Coulomb interaction of the form
$\frac{1}{2}U \sum_{\sigma \neq \sigma^{'}}  n_{\sigma}
n_{\sigma{'}}$, where $n_\sigma$ is the $f$-orbital (or
$d$-orbital) occupancy. For Fe, (i) delocalized $s, p$ electrons,
and (ii) localized $3d$ electrons.

\vspace{4 mm}
\noindent
The Hamiltonian for the spin orbitally degenerate systems is of the form
\begin{eqnarray}
\widehat{H} =
\sum_{i,j} \sum_{m, m'} \sum_{\sigma} t_{ij}^{mm'} \hat{c}^+_{im\sigma}
\hat{c}_{im'\sigma}
& + & \frac{(U-J)}{2} \sum_i \sum_{m \neq m'} \sum_{\sigma} \hat{n}_{im\sigma}
\hat{n}_{im'\sigma}
\nonumber \\
& + & \frac{U}{2} \sum_{i,m,m'} \sum_{\sigma}
\hat{n}_{im\sigma} \hat{n}_{im'-\sigma}
\end{eqnarray}
where $\hat{c}_{jm'\sigma}$ is an electron annihilation operator with
an orbital index $m$ and spin $\sigma(=\alpha, \beta)$. In the lattice site
$i,t_{ij}^{mm'}$ are the hopping integrals and $\hat{n}_{im'-\sigma}$ is the
number operator of the $f$ (or $d$) electron at site $i$, orbital $m$ with spin
$\sigma$. The first term in Eq. (5) describes the hopping of electrons
between lattice sites $i$ and $j$; the interactions between the localized
electrons are described by the second and third terms, where $U$ and
$J$ represent the on-site Coulomb and exchange interaction, respectively.

\vspace{4 mm}
\noindent
If we want to correct the LSDA functional for localized electrons we must
first extract their LSDA treatment to avoid double count of the interaction.
The spin density functional theory assumes a local exchange-correlation
potential which is a function of the local charge and spin densities, so,
fluctuations around the average occupations are neglected. In the mean field
approximation (MFA) we can write
\begin{eqnarray}
\hat{n}_{m \sigma} \hat{n}_{m' \sigma'} = \hat{n}_{m \sigma} n_{m'
\sigma'} + \hat{n}_{m' \sigma'} n_{m \sigma} - n_{m \sigma} n_{m'
\sigma'}
\end{eqnarray}
where $n_{m \sigma}$ is the mean value of $\hat{n}_{m \sigma}$ and
$n_{\sigma} = \sum_{m} n_{m \sigma}$. By introducing this approximation in
Eq. (5), we obtain the expression for potential energy in the mean
field approximation,
\begin{eqnarray}
E^{MF} = \frac{U-J}{2} \sum_i \sum_{\sigma} n_{i \sigma}
(n_{i\sigma} - n_{im\sigma}) + \frac{U}{2} \sum_i \sum_{\sigma}
n_{i \sigma} n_{i-\sigma}
\end{eqnarray}

\vspace{4 mm}
\noindent
Solovyev {\it et al.} [23] propose to extract an energy function from the
total number of electrons per spin $n_{i \sigma}$ which would act as the LSDA
potential. Such expression can be obtained from Eq.(5) in the atomic limit
where occupation of the individual particle $n_{i \sigma}$ is either 0 or 1:
\begin{eqnarray}
E_{cor} ^{LSDA} = \frac{U-J}{2} \sum_{i \sigma} n_{i \sigma} (n_{i
\sigma} -1) + \frac{U}{2} \sum_{i \sigma} n_{i \sigma}
n_{i-\sigma}
\end{eqnarray}
This energy is now subtracted from $E^{MF}$ to obtain that associated with
the total energy for localized states:
\begin{eqnarray}
\Delta E = E^{MF} - E_{cor}^{LSDA}
& = & \frac{U-J}{2} \sum_{i \sigma} n_{i \sigma} (1-n_{im \sigma})
\nonumber \\
& = & \frac{U-J}{2} \sum_{im \sigma} (n_{im \sigma} - n_{im \sigma}^2).
\end{eqnarray}

\vspace{4 mm}
\noindent
The fraction of the potential acting on the localized orbital $(m\sigma)$ is
found by differentiating Eq. (9) with respect to the occupation number
$n_{im\sigma}$:
\begin{eqnarray}
\frac{d \Delta E}{d(n_{i m \sigma})}= \Delta V_{i m
\sigma}=(U-J) \left ( \frac{1}{2} -n_{i m \sigma} \right ).
\end{eqnarray}
An orbital dependent one-electron potential is thus obtained.

\section{Results and discussion.}

\subsection{First Stage: LMTO-ASA calculations}

\vspace{4 mm} \noindent Results obtained in the optimization of
the intermetallic compound CeFeGe$_3$ are shown in Table II. The
LSD and GGS approximations yield values below the experimental
parameters, which may be due to the fact that neither
approximation removes the error introduced by the self-energy
interaction term, arising from double count in the Hamiltonian.
Furthermore, this material does not present a trend similar to
that of semiconductors since otherwise, the LSDA and GGA results
would give values remaining below and above the experimental
parameters, respectively. A similar situation is found in the
equilibrium properties of the plutonium (phase $\delta$-Pu)[24],
in which, results obtained with LDA and GGS give numbers that
remain below the corresponding experimental values. We work with
the CASTEP program at GGS level employing 446 K-points in the
Brillouin zone. On the other hand, the density of states (total
and partial) and band structure were obtained via the LMTO-ASA
methodology within the DFT theory and the LSDA approximation,
using the von Barth-Hedin parametrization at the optimized unit
cell geometry. The corresponding results are compiled in Table I.

\vspace{4 mm}
\noindent
The total density of states (DOS) is given in Fig. 1, whereas the partial DOS
associated with cerium $f$-states and iron $d$-states is illustrated in Figs.
2 and 3, respectively. They all indicate a metallic behavior for the material
here analyzed, since we have the Fermi level slightly displaced from the
central band. This is also supported by looking at the plot of band states
(BS), as shown in Fig. 4, where a dense concentration of bands due to the
cerium $f$ states occurs around the Fermi level. Furthermore, the fact
of having almost horizontal bands points to the characteristic behavior of a
heavy fermion compound (in our case, $\gamma=150$ mJ-molK$^2$) [1,5].

\vspace{4 mm} \noindent In the literature it is reported that the
greater magnetic contribution stems from cerium (being the
magnetic impurity responsible for the Kondo effect); a fact that
is corroborated in the present theoretical calculations (see Table
I). Once optimized the geometrical parameters, the material here
analyzed turns out to be one of those classified as an
intermediate valence compound. In our calculation, Ce has a
magnetic moment of $-$0.00132 $\mu_{\beta}$ whereas the compound's
total magnetic moment is $-$0.0010084 $\mu_{\beta}$. Therefore,
according to the criterion proposed by Vildosola and Llois [25],
i.e., {\sl by using calculations at the LSDA level and the
exchange-correlation potential of Perdew-Wang, they propose that
the materials can be classified as follows:}
\begin{eqnarray}
\fbox{$\begin{array} {rcl}
                  Itinerant\,\,\,if &\mu_{Ce} & = \,\,0,         \\
    Intermediate\,\,valence\,\,\,if &\mu_{Ce} & < \,\,0.5 \mu_B, \\
                   Magnetic\,\,\,if &\mu_{Ce} & > \,\,0.5 \mu_B
      \end{array}$
     }
\end{eqnarray}
An intermediate valence system is usually defined in the
literature as one having a noninteger average number of $f$
electrons. Note that the charge on the Ce atom according with the
LSDA calculation is of $-$2.4654 a.u. with a magnetic moment of
$-$0.00132. This means that the normal atom configuration of
$[Xe]4f^15d^16s^2$ is being added with almost 3 electrons whose
alpha and beta spins pair off one another, which results in an
overall moment of nearly zero magnitude.

\subsection{Second Stage: LSDA+U calculations}

\vspace{4 mm} \noindent As can be inferred from the analysis
presented in the previous stage, the DFT theory$-$because of its
intrinsic formulation$-$it cannot deal properly with strongly
correlated systems, therefore, we have chosen to resort to the
LSDA+U$^{SIC}$ approximation, developed by Anisimov $et$ $al.$
[9], to calculate the electronic structure of the intermetallic
compound $CeFeGe_3$, as implemented in the Stuttgart's TB-LMTO-ASA
code version 47. For this calculation, we employed the parameters
reported in the literature: $U=5.4$ eV [11] for cerium and $U=2.3$
eV (J = 0.9 eV) [12] for iron.

\vspace{4 mm}
\noindent
The resulting density of states in this case is plotted in Fig. 5, whereas
the partial DOS for cerium $4f$-states and iron $3d$-states are depicted in
Figs. 6 and 7, respectively. By virtue of the approximation used in this stage,
the latter include effects arising from strong correlation, not accounted for
in those obtained by means of a conventional DFT calculation.
The corresponding band structure is displayed in Fig. 8.

\vspace{4 mm} \noindent In the display of total DOS the greatest
contribution comes from the cerium $4f$-states where an increased
splitting within the energy bands$-$in addition to a relative
overall shift around the Fermi level$-$can be seen in Fig. 5. This
feature also shows up in the partial density of states, where a
large band energy splitting arising from the cerium $4f$-states
can be appreciated in Fig. 6 (being 5 eV wide, approximately), as
compared with the corresponding energy band display (Fig. 2)
obtained in the previous stage, in which, the presence of a nearly
single peak dominates, located very close and above the Fermi
level. The magnetic contributions to this material come partly
from both cerium and iron, even prior to the introduction of the
exchange parameter (previous stage), whereas the magnetic moment
for germanium is almost null, as seen in Table I. Furthermore, the
material presents metallic behavior. When carrying out
calculations with an exchange parameter $J$ equal to 0.90 eV for
iron there occurs a reduction of the magnetic moment in a $42\%$
rate, which probably follows from a slight localization effect of
electrons on the iron $3d$-shell. On the other hand, the high
concentration of energy bands occurring around the Fermi level is
consistent with a metallic behavior. In fact, a greater density of
bands is observed in this stage (see Fig. 8) as compared to that
observed without introducing the Coulomb parameter (see Fig. 4).
Such feature$-$in this stage$-$points to a typical characteristic
of heavy fermion materials.

\vspace{4 mm} \noindent According to the criterion mentioned in
the first stage, proposed by Vildosola and Llois [25], and
extended to handle the total magnetic moment per unit cell, one
can notice that the magnetic moments for cerium, $-1.1407
\mu_{\beta}$, iron, $(0.6147 \mu_{\beta})$, and $-0.00055
\mu_{\beta}$ for the empty spheres, give altogether a total moment
of $-0.552 \mu_{\beta}$. This falls within the classification
corresponding to an magnetic material. Also, the charge
distribution shown in Table I (empty spheres introduced by the ASA
condition), points to a possible covalent character among the
atoms of cerium, iron and germanium, although this manifests iself
very weakly, despite the fact of accounting for correlation
effects in the theoretical calculation. Comparing the charge of
the Ce atom in the LSDA(vBH) and LSDA(vBH)+U cases, it is seen to
be similar for both: $-$2.4654 a.u. and $-$2.3847 a.u.,
respectively. However, the $U$ interaction gives rise to a
complete change in the spin behavior: whereas in the absence of
$U$ almost 3 electrons are added to the neutral atom$-$although
the overall magnetic moment is nearly zero$-$now the salient
effect in the presence of the $U$ interaction is to align the
electrons, yielding an effective magnetic moment whose magnitude
is $-1.1407 \mu_{\beta}$.

\vspace{4 mm}
\noindent
Having no charge difference in the two cases, where the $U$ interaction acts
only on the $f$ electrons, the filling of the $f$-shell proceeds via splitting
of the ${\alpha}$ and ${\beta}$ levels by the energy $U$ and the occurrence of
electron alignment on the $f$-shell only. The remaining charge distributes
over the other shells: $s$, $p$ and $d$.

\section{Conclusions}

\vspace{4 mm}
\noindent
The calculations performed by means of the LMTO-ASA approximation, within the
DFT theory, lead to results that are similar to those reported in the
literature, in particular, the obtained magnetic contribution of the compound
here analyzed, which practically corresponds to that of a nonmagnetic material.
On the other hand, the partial DOS, together with the band structure, show a
metallic behavior for the latter. Furthermore, results obtained with
calculations carried out where the Coulomb parameters $U$ and $J$ (for iron)
are introduced, also favor a metallic behavior and, in addition, a heavy
fermion character for this material. The two-stage analysis performed in the
present study also indicates a small charge covalent character. The pronounced
magnetic moment reduction occurring in iron is here ascribed to an electronic
cloud localization on the $3d$-shell of this atom, which arises as a direct
consequence of taking into account strong electronic correlation effects.

\vspace{10 mm}
{\Large
\centerline{{\bf Acknowledgments}}}

\vspace{6 mm}
\noindent
Contract grant sponsors:  Consejo Nacional de Ciencia y Tecnolog\'{i}a
(CONACYT, M\'exico) and Vicerrector\'{i}a de Investigaci\'on y Estudios de
Posgrado (VIEP, M\'exico) at Benem\'erita Universidad Aut\'onoma de Puebla.
Contract grant numbers: 32213-E (CONACYT) and II-101I02 (VIEP).


\newpage
\vspace{10 mm} {\Large \centerline{{\bf Table and figure
captions}}}

\vspace{6 mm} \noindent {\bf Table I:} Parameters utilized in the
electronic structure calculations of the ternary compound
CeFeGe$_3$ and theoretical data obtained by using the TB-LMTO-ASA
approach.


\vspace{2 mm} \noindent {\bf Table II:} Cell parameters optimized
by means of the CASTEP program.

\vspace{2 mm} \noindent {\bf Fig. 1:} Total density of states of
CeFeGe$_3$ obtained by the LSDA-vBH approximation.

\vspace{2 mm} \noindent {\bf Fig. 2:} Partial density of states
(Ce 4$f$-states) obtained by the LSDA-vBH approximation.

\vspace{2 mm} \noindent {\bf Fig. 3:} Partial density of states
(Fe 3$d$-states) obtained by the LSDA-vBH approximation.

\vspace{2 mm} \noindent {\bf Fig. 4:} Band structure obtained by
the LSDA-vBH approximation.

\vspace{2 mm} \noindent {\bf Fig. 5:} Total density of states
obtained by the LSDA(vBH)+U approximation using the parameter U
for cerium and parameters U and J (exchange) for iron.

\vspace{2 mm} \noindent {\bf Fig. 6:} Partial density of states
(Ce 4$f$-states) obtained by the LSDA(vBH)+U approximation.

\vspace{2 mm} \noindent {\bf Fig. 7:} Partial density of states
(Fe 3$d$-states) obtained by the LSDA(vBH)+U approximation.

\vspace{2 mm} \noindent {\bf Fig. 8:} Band structure obtained by
the LSDA(vBH)+U approximation using the parameter U for cerium and
parameters U and J (exchange) for iron.

\newpage

 \centerline{{\bf Table I}}
\noindent \hspace {-1.5 cm}
\begin{tabular} {ccccccc} \hline
\noindent
                    &                         &            &    &    &    &  \\
\small Atoms in the & \small Crystallographic & \small MT sphere
                                                          & \small Charge$^a$ & \small Magnetic$^a$
                                                           & Charge$^b$ & Magnetic$^b$ \\
        CeFeGe$_3$ & Positions &  radii  & (a.u.) & Moment &  (a.u.)  & Moment  \\
              &         & (a.u.) &        & ($\mu_B$)&       & ($\mu_B$)\\
\hline
Ce   &(0.0, 0.0, 0.0)        &4.1077 &-2.4654 &-0.00132  &-2.3847 &-1.1407\\
Fe     &(1.0,1.0,0.66)         &2.4566 &-0.3027 &0.00033   &0.3339  &0.6147\\
Ge$_1$ &(0.5,0.0,0.25)         &2.5320 &1.1212  &-0.000005 &1.1208  &-0.01274\\
Ge$_2$ &(1.0,1.0,0.42)         &2.6284 &1.064   &-0.000013 &1.0499  &-0.02511\\
E$_1$* &(0.1224,0.1224,0.5684) &1.1281 &-0.1396 &-0.000001 &-0.1432 &-0.00055\\
E$_2$* &(0.1224,-0.1224,0.5684)&1.1281 &-0.1396 &-0.000001 &-0.1432 &-0.00055\\
E$_3$* &(-0.1224,0.1224,0.5684)&1.1281 &-0.1396 &-0.000001 &-0.1432 &-0.00055\\
E$_4$* &(-0.1224,-0.1224,0.5684)&1.1281&-0.1396 &-0.000001&-0.1432&-0.00055\\
\hline
\end{tabular}

\vspace{3 mm}{\small \noindent $^*$Positions of the empty spheres
that fulfill the
ASA condition within the LMTO-ASA approximation.\\
$^a$As obtained for the parameters optimized by means of  the
LSDA-vBH
\-approximation.\\
$^b$As obtanied via calculations performed with the LSDA(vBH)+U aproximation.}\\\\\\\\\\\\\\


\centerline{{\bf Table II}}

\vspace{2 mm} \noindent
\begin{tabular} {ccccc}
\hline
     &    &    &    &    \\
Experimental cell & Optimized cell & \% of error & Optimized cell
                                                  &\% of error  \\
parameters        & parameters     & Exp. vs LSDA & parameters
                                                  & Exp. vs GGS \\
  (\AA)           & (\AA) LSDA     &              & (\AA) GGS
                                                  &             \\
\hline
   a=b=4.332 & a=b=4.1767 & 3.71  & a=b=4.234  & 2.314  \\
     c=9.955 & c=9.5981   & 3.72   & c=9.73    & 2.312 \\
\hline
\end{tabular}

\end{sloppypar}

\vspace{8 mm}
\begin{thebibliography}{40}
\bibitem{} G. R. Stewart, \emph{ Rev. Mod. Phys.} {\bf 56}, 755 (1984); Peter Fulde, \emph{J. Phys.
F: Met. Phys.} {\bf 18}, 601 (1988).
\bibitem{} N. B. Brandt and V. V. Moshchalkov, \emph{ Adv. Phys.} {\bf 33}, 373 (1984).
\bibitem{} A. C. Hewson, \emph{The Kondo Problem to heavy Fermions} (Cambrige
University Press, 1997).
\bibitem{} Julian G. Sereni,  \emph{Rev. Esp. F{\'i}s.} {\bf 13} (1),  25  (1999).
\bibitem{} H. Yamamoto, H. Sawa and M. Ishikawa, \emph{Phys. Lett. A} {\bf 196}, 83 (1994);
H. Yamamoto, M. Ishikawa, K. Hasegawa
       and J. Sakurai, \emph{Phys. Rev. B.} {\bf 52}, 10136 (1995); E. Chigo Anota, J.
F. Rivas Silva, A. Bautista Hernandez and A.
       Flores Riveros, \emph{Superficies y Vac{\'i}o} {\bf 16} (1), 17 (2003).
\bibitem{} P. Fulde,  \emph{cond-mat/9803299}(1998); E. Chigo-Anota and J. F. Rivas-Silva, \emph{Superficies y
Vac{\'i}o} {\bf 17}, xxx (2004); E. Chigo-Anota and J. F.
Rivas-Silva, \emph{Rev. Soc. Qu{\'i}m. M{\'e}x.} {\bf 47} (3), 221
(2003).
\bibitem{} P. Hohenberg and W. Kohn, \emph{Phys. Rev. B} {\bf 136}, 864 (1964); W. Kohn and
L. J. Sham, \emph{Phys. Rev. A} {\bf 140}, 1133 (1965);
       W. Kohn, et al., \emph{J. Phys. Chem.} {\bf 100}, 12974 (1996); K. Capelle,
\emph{cond-mat/0211443} v1 November (2002).
\bibitem{} Hans L. Skriver, \emph{The LMTO Method} (Springer-Verlag, 1984); O. K.
Andersen and O. Jepsen, \emph{Phys. Rev. Lett.} {\bf 53},
       2571 (1984); O. Jepsen, G. Krier, A. Burkhardt and O. K. Andersen,
\emph{The TB-LMTO-ASA  program}, Max-Planck
        Institute, Stuttgart, Germany (1995).
\bibitem{} V. I. Anisimov, I. V. Solovyev, et al., \emph{Phys. Rev. B} {\bf 48}, 16929
(1993); A. I. Liechtenstein, V. I. Anisimov, and J.
       Zaanen, \emph{Phys. Rev. B} {\bf 52}, 5467 (1995); V. I. Anisimov, F.
Aryasetiawan, and A. I. Liechtenstein, \emph{J. Phys.: Condens.
       Matter} {\bf 9}, 767 (1997); A. B.  Shick, A. I. Liechtenstein and W. E.
Pickett, \emph{Phys. Rev. B} {\bf 60}, 10763 (1999);
     E. Chigo Anota  and  J.  F. Rivas-Silva, \emph{Rev. M{\'e}x.
F{\'i}s.} {\bf 50}, xxx (2004).
\bibitem{} Cerius$^2$ Versi{\'o}n 4.2 MatSci, \emph{Manual of CASTEP}, Molecular  Simulations
Inc. (2000).
\bibitem{} B. E. Min, H. J. F. Jansen, T. Oguchi and A. J. Freeman, \emph{Phys. Rev. B}
{\bf 33}, 8005 (1986).
\bibitem{} A. M. Oles and G. Stolhoff, \emph{Phys. Rev. B} {\bf 29}, 314 (1984).
\bibitem{} U. von Barth and L. Hedin, \emph{J. Phys. C } {\bf 5}, 1629 (1972) .
\bibitem{} D. M. Ceperly and B. J. Alder, \emph{Phys. Rev. Lett.} {\bf 45}, 566 (1980).
\bibitem{} D. C. Langreth and M. J. Mehl, \emph{Phys. Rev. Lett.} {\bf 47}, 446 (1981);  C. D.
Hu and D. C. Langreth, \emph{Physica Scripta} {\bf 32}, 391
       (1985).
\bibitem{} J. P. Perdew and Y. Wang, \emph{Phys Rev. B} {\bf 33}, 8800 (1986).
\bibitem{} D. Vanderbilt, \emph{Phys. Rev. B}  {\bf 41}, 7892 (1990).
\bibitem{} L. Klienman and D. M. Bylander, \emph{Phys. Rev. Lett.} {\bf 48}, 1425 (1982).
\bibitem{} J. P. Perdew and A. Zunger, \emph{Phys. Rev. B} {\bf 23}, 5048 (1981).
\bibitem{} J. P. Perdew, K. Burke and M. Ernzerhof, \emph{Phys, Rev. Lett.} {\bf 77}, 3865
(1996).
\bibitem{} H. J. Monkhorst and J. D. Pack, \emph{Phys. Rev. B}  {\bf 13}, 5188 (1976).
\bibitem{} O. H. Nielsen and R. M. Martin, \emph{ Phys. Rev. Lett.}  {\bf 28}, 697 (1983).
\bibitem{} I. V. Solovyev, P. H. Dederichs and V. I. Anisimov, \emph{Phys. Rev. B}
{\bf 50}, 16861 (1994).
\bibitem{} J. Boucher, B. Siberchicot, F. Jollet and A.  Pasturel, \emph{J. Phys.:
Condens Matter} {\bf 12}, 1723 (2000); D. Price, B. R.
        Cooper, S. P. Lim, I. Avgin, \emph{ Phys. Rev. B} {\bf 61}, 9867 (2000);  S. Y.
Savrasov and G. Kotliar,  \emph{ Phys. Rev. Lett.} {\bf 84}, 3670
        (2000).
\bibitem{} V. L. Vildosola and A. M. Llois, \emph{cond-mat/0001054}  v1 January (2000).
\end{thebibliography}
\end{document}